# Orientation selection and superconducting properties of epitaxial Al on ferromagnetic semiconductor (In,Fe)As

Hirotaka Hara[1], Keita Ishihara[1], Masaaki Tanaka[1,2,3,†], and Le Duc Anh[1,2,†]

[1] *Department of Electrical Engineering and Information System, The University of Tokyo, Tokyo, Japan*

[2] *Center for Spintronics Research Network, The University of Tokyo, Tokyo, Japan*

[3] *Institute for Nano Quantum Information Electronics, The University of Tokyo, Tokyo, Japan*

† Corresponding author: masaaki@ee.t.u-tokyo.ac.jp, anh@cryst.t.u-tokyo.ac.jp

**Abstract**

Superconductor/ferromagnet heterostructures provide a versatile platform for exploring spin-dependent superconducting phenomena arising from interfacial proximity effects. In this article, we investigate the structural and superconducting properties of Al thin films grown *in situ* by molecular beam epitaxy on strained Fe-doped ferromagnetic semiconductor (FMS) (In,Fe)As layers. X-ray diffraction and transmission electron microscopy reveal the epitaxial growth of single-crystalline Al layers, with the growth orientation changing from (110) to (111) as the in-plane lattice constant of (In,Fe)As increases. The superconducting critical temperature of Al varies systematically with the film surface morphology and grain size. In addition, the critical magnetic field of Al exhibits an anomalous decrease below 0.5 K, possibly reflecting magnetic coupling to the underlying FMS (In,Fe)As layer. These findings provide a guideline for material design of epitaxial Al/(In,Fe)As heterostructures, which may serve as a promising platform for investigating proximity-induced superconducting and magnetic phenomena in semiconductor-based hybrid quantum devices.

**Main text**

Heterostructures composed of superconductors and ferromagnetic materials offer versatile platforms for exploring a wide range of physical phenomena. These include oscillations of the superconducting order parameter in a ferromagnet, which are often discussed in relation to Fulde–Ferrell–Larkin–Ovchinnikov (FFLO) states [1,2], 0-π Josephson junctions, and emergence of spin-triplet superconductivity [3,4,5,6,7]. To date, ferromagnetic metals and ferromagnetic insulators have been predominantly employed in contact with superconductors [5]. In comparison with these conventional systems, heterostructures combining superconductors with ferromagnetic semiconductors (FMSs) [6,7] offer an important advantage: The transport and magnetic properties of the FMS channel can be tuned electrically by applying a gate voltage. This tunability enables control over the density, phase, and spin angular momentum of Cooper pairs, thereby opening avenues to novel quantum phenomena and devices, including gate-tunable 0-π Josephson junctions [5], superconducting diodes [8], and potential platforms for topological superconductivity and Majorana physics [9,10].

For the formation of high-quality, transparent interfaces with superconductors, Fe-doped III-V FMSs are a particularly promising materials platform [11]. These materials are formed by doping III-V semiconductors such as InAs [12], GaSb [13], InSb [14] and AlSb [15], with several percents of Fe. Fe-doped FMSs largely preserve the crystal and electronic band structures of the host semiconductors while exhibiting ferromagnetism. Among them, (In,Fe)As is especially attractive because it inherits the structural and electronic properties of n-type InAs. In (In,Fe)As, conduction electrons occupy states near the conduction band minimum, where they have a small effective mass and long coherence length [16,17,18]. In addition, Fermi-level pinning above the conduction-band minimum at the surface is expected to enable Ohmic contacts with metal electrodes. Therefore, highly transparent interfaces can be formed between superconducting metals such as Al and (In,Fe)As, analogous to the case of Al/InAs hybrid structures [19,20,21,22]. Furthermore, (In,Fe)As demonstrates carrier-induced ferromagnetism which can be controlled by electrostatic gating [17]. It also exhibits a large spontaneous spin splitting in the conduction band [18,23] and a capability for ultrafast magnetization modulation [24]. Spin-triplet superconductivity has also been observed in Josephson junctions based on Nb/(In,Fe)As/Nb structures [6,7]. However, in those studies, the Nb/(In,Fe)As bilayers were fabricated *ex situ*, with Nb films sputtered onto a molecular beam epitaxy (MBE)-grown (In,Fe)As thin film, which can degrade the quality of the superconductor-semiconductor interface. These results underscore the necessity for further endeavors to achieve high-quality Al/(In,Fe)As hybrid thin film structures using *in situ* deposition. Such epitaxial thin-film heterostructures combine clean superconductor–semiconductor interfaces with compatibility for planar device processing, making them promising building blocks for scalable superconducting spintronic and hybrid quantum devices.

In this work, we investigated the structural and superconducting properties of Al thin films epitaxially grown on (In,Fe)As using MBE. The growth of Al on zinc-blende III-V semiconductors is known to be strongly influenced by the lattice constant of the underlying semiconductor [25,26]. We therefore grow (In,Fe)As films on ternary $(In_{1-y},Al_y)As$ buffer layers on InP (001) substrates, which impose tunable compressive epitaxial strain on the FMS layer. Varying the Al composition $y$ in the $(In_{1-y},Al_y)As$ layer has been demonstrated to result in a wide range of control of structural, transport and magnetic properties of the (In,Fe)As thin films [27]. By investigating the properties of Al thin films epitaxially grown on (In,Fe)As under these varied conditions, we aim to obtain useful insights into fabrication strategies and material design for Al/Fe-doped FMS heterostructures.

We prepared three heterostructures consisting of 20-nm Al / 15-nm (In,Fe)As / $(In_{1-y},Al_y)As$ graded buffer layers on InP (001) substrates, as shown in Fig. 1(a). The (In,Fe)As layers contain 9.6% Fe, and the Al compositions at the top of the $(In_{1-y},Al_y)As$ buffer layers were $y_{top}$ = 0.10, 0.15, and 0.19. These samples are denoted #10, #15, and #19, respectively, and were designed to impose different magnitudes of in-plane compressive strain, $\varepsilon_{in\text{-}plane}$, on the (In,Fe)As layers. The (In,Fe)As layers were grown by low-temperature MBE, and the growth details are described elsewhere [27]. After growth of the (In,Fe)As layers, the samples were transferred under ultrahigh vacuum to a metal-deposition chamber and cooled with liquid nitrogen for 90 min to reach a sufficiently low substrate temperature for the Al growth. Subsequently, 20-nm-thick Al layers were grown at $T_s \sim -60$°C with a growth rate of 1 nm/min.

The structural properties of the Al films were characterized by X-ray diffraction (XRD) and high-resolution transmission electron microscopy (TEM). Figure 1(b) shows $\theta$-$2\theta$ XRD scans, confirming that the heterostructures were grown as intended. The highest peaks at 63.4 degree and the relatively broad group of peaks between 61 and 63 degree are attributed to the InP substrate and the $(In_{1-y},Al_y)As$ graded buffer layer, respectively. As the Al composition $y_{top}$ in the buffer layer increases, the peak assigned to (In,Fe)As around 60-61 degree shifts to lower values of $2\theta$, indicating an increase in the out-of-plane (perpendicular) lattice constant of (In,Fe)As. Figure 1(c) summarizes the estimated in-plane lattice constant of (In,Fe)As and the corresponding compressive strain $\varepsilon_{in\text{-}plane}$, where the negative and positive signs denote compressive and tensile strain, respectively. Assuming biaxial strain in the epitaxial film, the in-plane lattice constant and strain were estimated from the measured out-of-plane lattice constant using the stress-strain relation and the elastic stiffness constants of InAs. These results confirm that the in-plane compressive strain in (In,Fe)As can be systematically controlled by varying the Al composition $y_{top}$ in the $(In_{1-y},Al_y)As$ graded buffer layer.

In addition, the inset of Fig. 1(b) shows that samples #10 and #15 exhibit a diffraction peak at 38.5 degree, which is assigned to Al (111) reflection, whereas no corresponding peak is observable for sample #19. For face-centered cubic Al, the Al (220) reflection is expected to appear near 65.4 deg. Although peaks around 65 degree are observable for all samples in Fig. 1(b), they remain after wet etching of the Al layers and are also observed in samples without Al growth, as reported previously [27]. This indicates that the peaks near 65 degree originate from the buffer layers rather than from Al. Therefore, among the observed diffraction features, only the peak at 38.5 degree can be unambiguously assigned to Al. Its presence in sample #10 and #15 and absence in #19 suggest that the dominant growth orientation of Al changes depending on the lattice constant of the underlying (In,Fe)As layer.

We further examined the structural evolution of Al using TEM. Figures 1(d) and 1(e) show TEM images of the upper regions of samples #10 and #19, respectively. Domain structures accompanied by Moiré patterns are observed in the Al layers. Higher-magnification TEM images confirm that the semiconductor layers up to the (In,Fe)As layer retain the zinc-blende crystal structure, whereas the Al layers consist of single-crystalline domains. The Al growth orientation is identified as (111) in sample #10 and (110) in sample #19, as shown in Figs. 1(f) and 1(g), respectively. In sample #10, the Moiré pattern originates from interference between (111)-oriented Al domains with in-plane rotational mismatch, as confirmed by transmission electron diffraction (TED). These observations indicate that the dominant growth orientation of Al is (111) in sample #10 and (110) in sample #19, consistent with the XRD results. The relationship between the Al growth orientation and the lattice constant of (In,Fe)As (or equivalently the compressive strain), is summarized in Table 1.

The MBE growth of Al on zinc-blende III-V semiconductors has been studied for several decades in the context of electronic and quantum-device applications. For semiconductors with lattice-constants near 6.0Å, Wang *et al*. reported that Al grows along the [111] ([110]) direction when the lattice constant of the underlying semiconductor is smaller (larger) than 5.98 Å [26]. However, this trend is opposite to our observation. In our samples, the dominant Al growth orientation changes from (110) to (111) as the in-plane lattice constant of (In,Fe)As increases from 5.99 to 6.03 Å, indicating that the lattice constant alone does not fully determine the Al growth orientation. Another important factor is the chemical composition of the topmost semiconductor surface. When Al is deposited directly on InAs, the resulting film can exhibit roughness larger than the nominal film thickness because of the interfacial reaction between Al and InAs [28]. In contrast, the insertion of a thin semiconductor layer, such as GaAs or AlAs with a thickness ranging from subnanometer to a few nanometers, can yield flat Al films with growth orientations that depend on the inserted material [29,30]. Such material-dependent effects have also been observed when the underlying semiconductor is a ternary alloy, such as (In,Ga)As [30,31]. In our samples, the Fe atoms incorporated into (In,Fe)As may play a surfactant-like role, suppressing interfacial roughening even when Al is deposited directly on the (In,Fe)As surface. They may also influence the Al growth orientation, potentially modifying the relationship between the preferred Al orientation and the lattice constant of the underlying semiconductor from that reported in the previous study [26]. In addition, an In-rich surface termination may be present prior to Al deposition, in contrast to the As-rich surfaces typically obtained during conventional III–V semiconductor growth, because the (In,Fe)As layers were grown by low-temperature molecular beam epitaxy. Such differences in surface termination can modify the chemical bonding, atomic arrangement, and nucleation energetics at the Al/semiconductor interface, thereby changing the relative stability of different Al growth orientations [32,33]. Therefore, the observed transition in the dominant Al growth orientation from (110) to (111) with increasing in-plane lattice constant of (In,Fe)As may arise from a combination of lattice-constant effects and the surface chemistry of (In,Fe)As.

Next, we characterized the superconducting properties of the Al layers by four-terminal transport measurements using cleaved samples. Figure 2(a) shows temperature dependence of the resistance for the three samples. All the samples exhibit zero resistance below 1.1 - 1.4 K upon cooling. The zero-resistance state is suppressed by both in-plane and perpendicular-to-plane magnetic field, and the corresponding behavior depends on temperature, as shown in Section 1 in Supplementary Material. From these data, the temperature dependence of the critical magnetic field $B_c$ was extracted, as shown in Figs. 2(b) and 2(c). In all samples, $B_c$ increase with decreasing temperature, and the data are well fitted by the equation $B_c(T) = B_c(0)\left[1 - \frac{e^{2\gamma}}{3}\left(\frac{T}{T_c}\right)^2\right]$ from the BCS theory [34], where $B_c(0)$ and $T_c$ are fitting parameters and $\gamma = 0.5772$ is the Euler constant. Two features are noteworthy in the temperature dependence of $B_c$: First, both $B_c$ and $T_c$ increase as the in-plane compressive strain in (In,Fe)As decreases, or equivalently as the in-plane lattice constant of (In,Fe)As increases. Second, $B_c$ begins to decrease as temperature decreases below 0.5 K for both in-plane and perpendicular magnetic fields.

One possible explanation for the systematic variation of $B_c$ and $T_c$ with the compressive strain in (In,Fe)As is a corresponding change in the magnetic properties of (In,Fe)As. However, previous measurements on samples with the same heterostructure showed that the saturation magnetization and coercive field of (In,Fe)As, which are approximately 40 emu/cm$^3$ and 2-5 mT, respectively, at 5 K, and do not vary substantially among the samples [27]. Therefore, the variation in the superconducting properties of Al is unlikely to originate directly from differences in the ferromagnetism of the underlying (In,Fe)As. Instead, the change in $T_c$ appears to be more closely related to the surface morphology of the Al layers. Figures 3(a)-3(c) show atomic force microscopy (AFM) images of the Al surfaces. The root-mean-square roughness and the average grain size of Al both increase as the in-plane compressive strain in (In,Fe)As increases from sample #10 to #19, as summarized in Table 2. In conventional BCS superconductors, including Al, $T_c$ is known to increase with decreasing the grain size [35,36]. This enhancement has been attributed either to the quantization of electronic levels contributing to superconductivity [37] or to modifications of the phonon spectrum caused by grain boundaries [38,39]. At grain boundaries, phonon amplitudes are enhanced because of reduced atomic symmetry. Garland and Watton [38,39] described the resulting increase in $T_c$ by the following expression:

$$\ln\left(\frac{T_c}{T_{c0}}\right) = -\frac{1}{2}\frac{k}{a} + \frac{1+\lambda}{A\left(1-\frac{1}{2}\mu^*\right)\lambda-\mu^*} - \frac{1+\lambda\left(1+\frac{k}{a}\right)}{A\left(1-\frac{1}{2}\mu^*\right)\lambda\left(1+\frac{k}{a}\right)\left[1-\frac{1}{2}\beta\left(\frac{k}{a}\right)\right]-\mu^*} \quad (1).$$

Here, $T_{c0}$ is the bulk critical temperature, $a$ is the grain size, $k$ is the surface enhancement of the electron-phonon coupling parameter, $\beta$ is the lattice-disorder parameter, $\lambda$ is the electron-phonon coupling constant, $A$ is a constant of order unity, and $\mu^*$ is the Coulomb pseudopotential. The relationship between $T_c$ and grain size $a$ reported by Pettit et al. was well reproduced by Eq. (1) [40]. Our data can also be described by this equation using the parameters reported in Ref. [40], as shown in Fig. 3(d). Although the mechanism by which the Al grain size is linked to the lattice constant of the underlying (In,Fe)As remains unclear, the observed variation in surface morphology may account for the systematic change in the superconducting properties of the Al layers.

On the other hand, the slight decrease in $B_c$ below 0.5 K may be related to enhanced ferromagnetic behavior in (In,Fe)As at low temperature. Figure 4(a) shows the magnetoresistance, defined as MR = $[R(B) - R(0)] / R(0)$ where $R$ is the resistance of (In,Fe)As measured under perpendicular magnetic field at various temperatures. The detailed temperature dependence is provided in Section 2 in Supplementary Material. The sample used for the magnetoresistance measurement has the same structure as sample #10, but without the top Al layer. Meanwhile, magnetoresistance measurements for (In,Fe)As layers with compressive strains equivalent to those in samples #15 and #19 were difficult because of their insulating behavior at low temperature [27]. As shown in Fig. 4(a), the (In,Fe)As layer exhibits negative magnetoresistance with hysteresis over the entire temperature range ($T$ = 300 mK – 5 K), which is well below the Curie temperature 40 K of (In,Fe)As. In addition, the magnetoresistance peak near 0 T begins to split below 1 K, and pronounced peaks for which the resistance exceeds $R(0)$ appear at temperature below 1 K. The temperature dependence of the field at which the magnetoresistance peak appears, corresponding to the coercive field of (In,Fe)As, is summarized in Fig. 4(b). These results suggest that the magnetization process of (In,Fe)As changes markedly below 1 K, a behavior that has not been reported previously. Although such behavior is unusual for conventional ferromagnets, it may be related to a change in magnetic anisotropy, as observed in other FMSs [41,42]. The easy magnetization axis of the (In,Fe)As layer may be changed from in-plane ($T$ > 1.5 K) to perpendicular ($T$ < 1.5 K).

The ferromagnetism of (In,Fe)As could affect superconductivity in the Al layer in at least two ways: through stray magnetic field and through magnetic proximity effects, i.e., an induced exchange field at the interface. To assess the effect of stray field, we calculated the magnetic field generated by the (In,Fe)As layer. The calculation, the details of which are provided in Section 3 in Supplementary Material, indicates that the stray field emerging from (In,Fe)As is too small to account for the observed decrease in $B_c$. By contrast, an exchange field may provide a plausible explanation. In superconductor/ferromagnet bilayers, spin-dependent scattering at the interface can induce ferromagnetic order in the superconductor, known as the inverse (magnetic) proximity effect, leading to an exchange field that produces a spin-split density of states in a superconductor [43,44,45]. For example, in Al/EuS heterostructures, tunneling spectroscopy has revealed the exchange field of several tesla, depending on the external magnetic field, Al thickness, and temperature [46,47,48,49]. The exchange field can be theoretically expressed as

$$g\mu_{\mathrm{B}}H_{\mathrm{ex}} \propto \frac{G_{\Phi}^{\mathrm{s}}}{G_{\mathrm{s}}} \cdot \frac{\hbar D_{\mathrm{s}}}{d_{\mathrm{s}}^{2}}, \tag{2}$$

where $H_{ex}$, $G_{\Phi}^{s}$, $G_s$, $D_s$, $d_s$ denote the exchange field, spin-mixing conductance, normal-state conductance of the superconductor, diffusion constant, and thickness of the superconductor, respectively [50]. In our Al/(In,Fe)As system, a quantitative estimate of the exchange field is difficult because the spin-mixing conductance is highly sensitive to the interfacial properties. Nevertheless, the Al layers in Refs. [46,47] are only approximately one-eighth to one-half as thick as those in our samples. Considering the $1/d_s^2$ dependence, the exchange field in our case could be substantially reduced, possibly to the range of several tens to hundreds of milliTesla, if the other parameters are comparable. In addition, because (In,Fe)As is a diluted magnetic material and the theoretical maximum local magnetic moment is 5 $\mu_B$ per Fe, smaller than 7 $\mu_B$ per Eu in EuS, the spin-mixing conductance in Al/(In,Fe)As may also be smaller. This reduction could be consistent with the range from several-mT to several tens of mT of the decrease in $B_c$ observed at $T < 0.5$ K in Figs. 2(b) and 2(c). Although this interpretation remains semi-quantitative, the possible signature of an exchange field suggests that the Al/(In,Fe)As interface may host the type of magnetic interaction expected in superconductor/ferromagnet heterostructures. This interaction is an important prerequisite for several spin-dependent superconducting phenomena. Further theoretical and experimental investigations are needed to clarify the role of the exchange field in this system.

**Conclusion**

In this study, we investigated heterostructures composed of superconducting Al and the Fe-doped FMS (In,Fe)As grown by MBE. The Al layers were deposited on (In,Fe)As with different in-plane lattice constants, which were controlled by epitaxial compressive strain imposed by the underlying (In,Al)As buffer layers. As the in-plane lattice constant of (In,Fe)As increases, the dominant Al growth orientation changes from (110) to (111). This behavior may also be influenced by the presence of Fe in (In,Fe)As.

The Al layers exhibit superconductivity even when grown directly on (In,Fe)As, and their critical temperature $T_c$ and critical magnetic field $B_c$ were evaluated. The dependence of $T_c$ on the lattice constant of the underlying (In,Fe)As can be explained in terms of the grain size of the Al layers, consistent with previous studies. The decrease in $B_c$ below 0.5 K may be related to the enhancement of magnetization in (In,Fe)As, which could induce a magnetic proximity effect to the superconducting Al layer. These findings provide a guiding insight into the material design of Al/(In,Fe)As heterostructures, which are a promising materials platform for investigating interfacial coupling between superconductivity and ferromagnetism in semiconductor-based hybrid quantum devices.

**Supplementary Material**

The Supplementary Material provides 1. magnetic field dependence of the resistance of Al in the low-temperature regime where it is superconducting, 2. magnetoresistance of (In,Fe)As from room temperature down to cryogenic temperatures, and 3. estimated stray field at the position of the Al layer generated by the adjacent (In,Fe)As layer.

**Acknowledgements**

This work was supported in part by the Grants-in-Aid for Scientific Research (19K21961, 20H05650, 22K18293, 23K17324, 24H00018, 25H00840 and 25K24621), the CREST program (JPMJCR1777), NEXUS (JPMJNX25D5) program of JST, and the Spintronics Research Network of Japan (Spin-RNJ). This work was partly performed using facilities of the Cryogenic Research Center, the University of Tokyo. This work was supported by ARIM of MEXT (JPMXP1225UT0058).

**Author declarations**

**Conflict of interest**

The authors declare no competing interests.

**Author contributions**

H.H. and K.I. grew samples, and H.H. conducted measurements. H.H. prepared initial drafts. L.D.A and M.T. supervised the experiments, contributed data discussions, and revised the manuscript. All authors discussed the results.

**Data availability**

The main data that support the findings of this study are available in this article and its Supplementary Material. Additional data are available from the corresponding author upon request.

## Tables and Figures

**Table 1.** Relationship among the Al composition $y_{top}$ in the $(In_{1-y},Al_y)As$ buffer layer, the in-plane lattice constant of (In,Fe)As, $a_{(In,Fe)As,\ in\text{-}plane}$, the in-plane strain $\varepsilon_{in\text{-}plane}$ applied to (In,Fe)As, and the dominant growth orientation of Al. The Al composition $y_{top}$ also corresponds to the sample name.

| Sample | Al ratio $y_{top}$ in the $(In_{1-y},Al_y)As$ | $a_{(In,Fe)As,\ in\text{-}plane}$ (Å) | $\varepsilon_{in\text{-}plane}$ (%) | Al growth orientation |
|---|---|---|---|---|
| #10 | 0.10 | 6.033 | -0.42 | (111) |
| #15 | 0.15 | 6.010 | -0.79 | (111) |
| #19 | 0.19 | 5.994 | -1.06 | (110) |

**Table 2.** Relationship between the surface morphology of Al and its superconducting critical temperature $T_c$. The surface-morphology parameters include the root mean square (RMS) roughness and the average grain size, which was obtained from Fourier analysis of the AFM images shown in Figs. 3(a)-3(c). $T_c$ was estimated from the BCS fitting shown in Fig. 2(b).

| Sample | RMS roughness (nm) | Average grain size (Å) | $T_c$ (K) |
|---|---|---|---|
| #10 | 0.5249 | 340 | 1.43 |
| #15 | 0.6014 | 500 | 1.26 |
| #19 | 0.7245 | 590 | 1.19 |

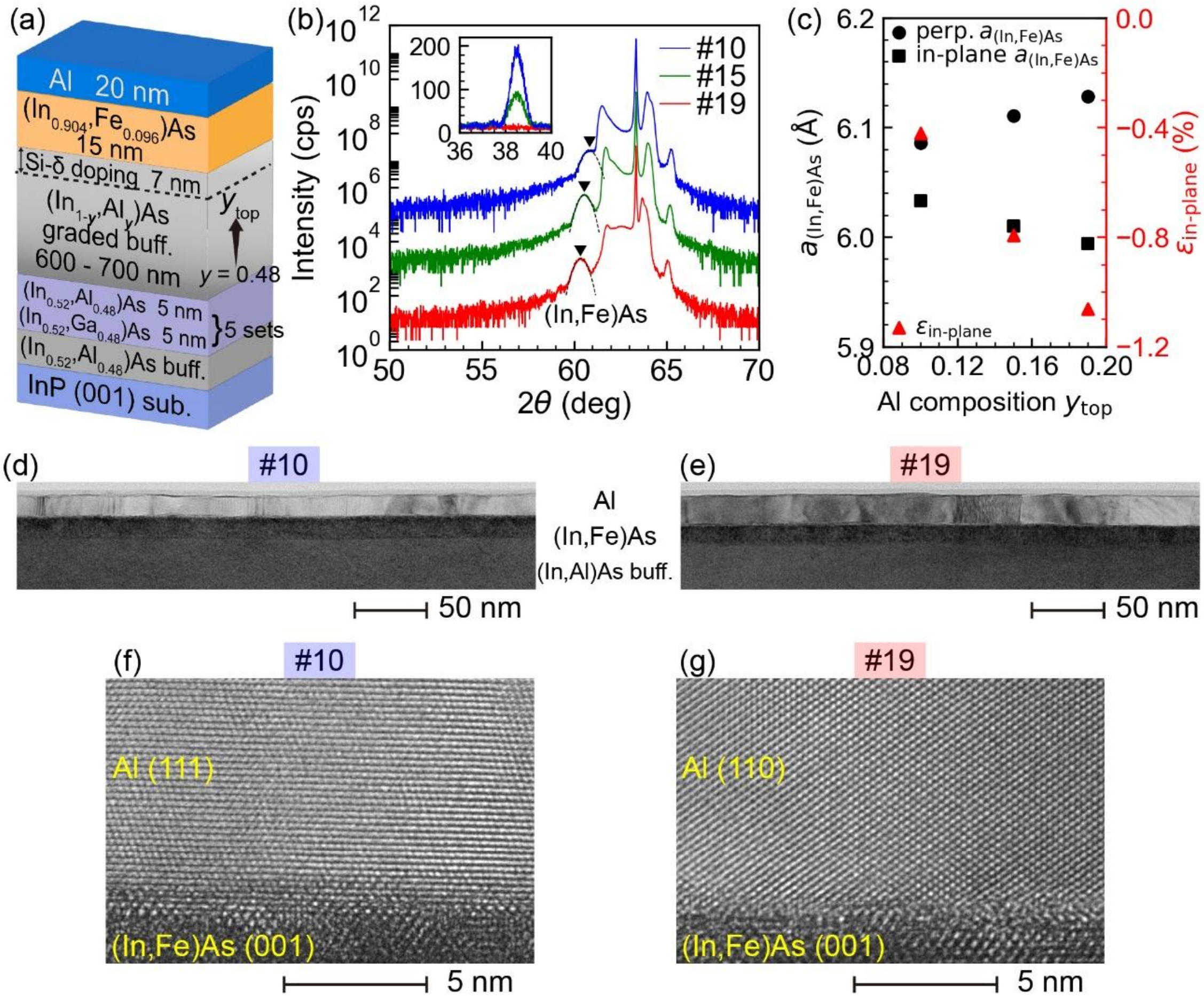

**Figure 1.** (a) Schematic illustration of the heterostructure examined in this study. (b) XRD $\theta$-$2\theta$ scans over the $2\theta$ range from 50 to 70 degree. For clarity, the data for samples #10 and #15 are vertically offset by 10000 and 100, respectively. Black triangles indicate the peaks from (In,Fe)As, and the black dashed curves represent Gaussian fits to these peaks. The inset shows scans over the $2\theta$ range from 36 to 40 degree. The peaks around 38.5 degree are attributed to the face-centered cubic Al (111) reflection. (c) Out-of-plane (perpendicular) lattice constant of (In,Fe)As obtained from Fig. 1(b), together with the estimated in-plane lattice constant and compressive strain of (In,Fe)As. (d), (e) TEM images of samples #10 and #19, respectively. From top to bottom, the images show the Al layers, the dark (In,Fe)As layers, and the (In,Al)As buffer layers. (f), (g) High-resolution TEM images of samples #10 and #19, respectively. The incident electron beam directions in these TEM images is along the [110] axis of the InP (001) substrates.

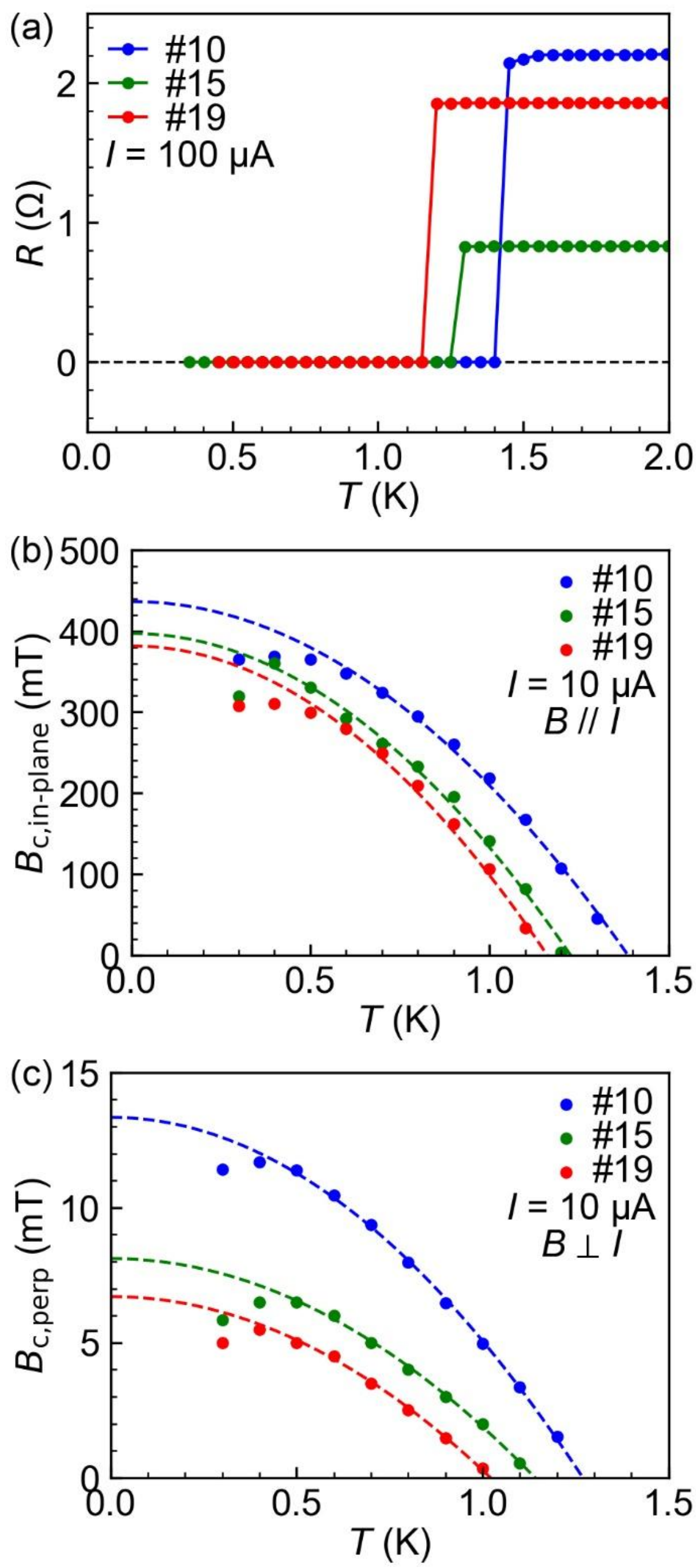

**Figure 2.** (a) Temperature dependence of the resistance of the Al layers measured with a bias current of 100 μA. (b), (c) Temperature dependence of the critical magnetic field $B_c$ of the Al layers under in-plane and perpendicular magnetic fields, respectively. The dashed curves are fitting curves based on the BCS theory. The bias current is 10 μA.

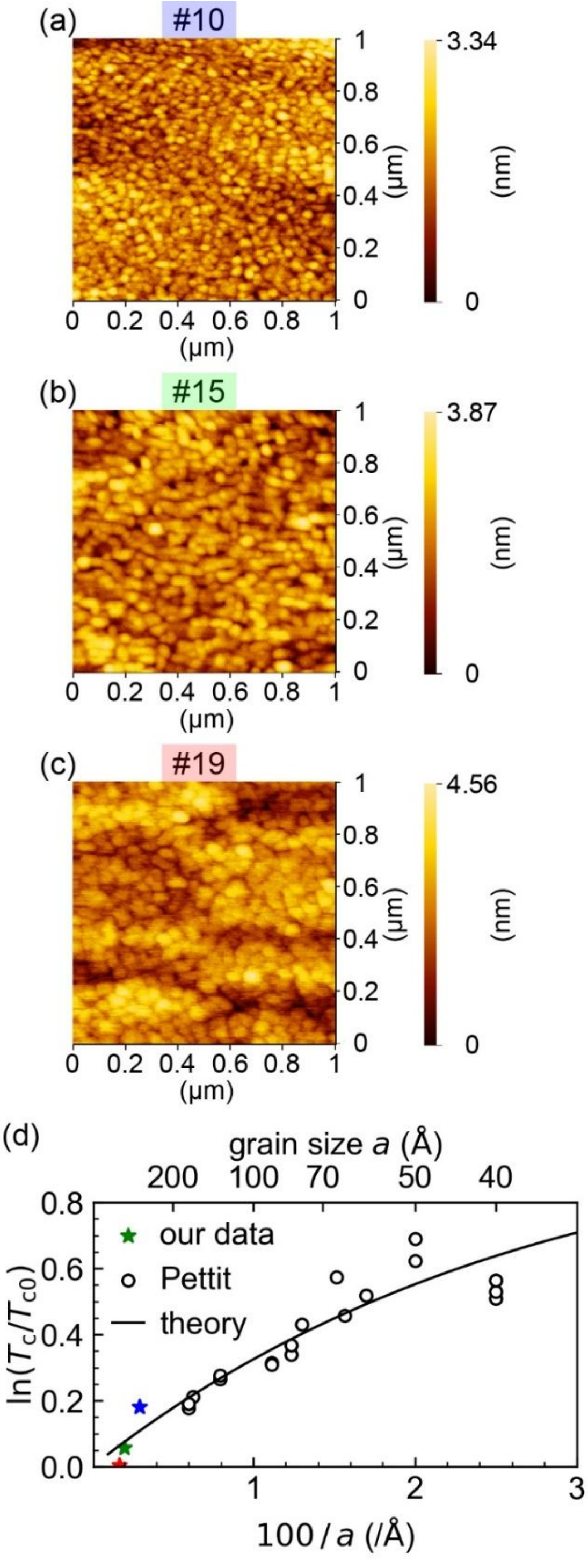

**Figure 3.** (a)-(c) AFM images of the Al layers. (d) Relationship between the enhancement of the superconducting critical temperature $T_c$ relative to the bulk value $T_{c0}$, and average grain size $a$. The colored stars represent our data, whereas the open circles denote the data reported by Pettit [36]. The solid line shows the theoretical curve from the works of Garland and Watton [34,35].

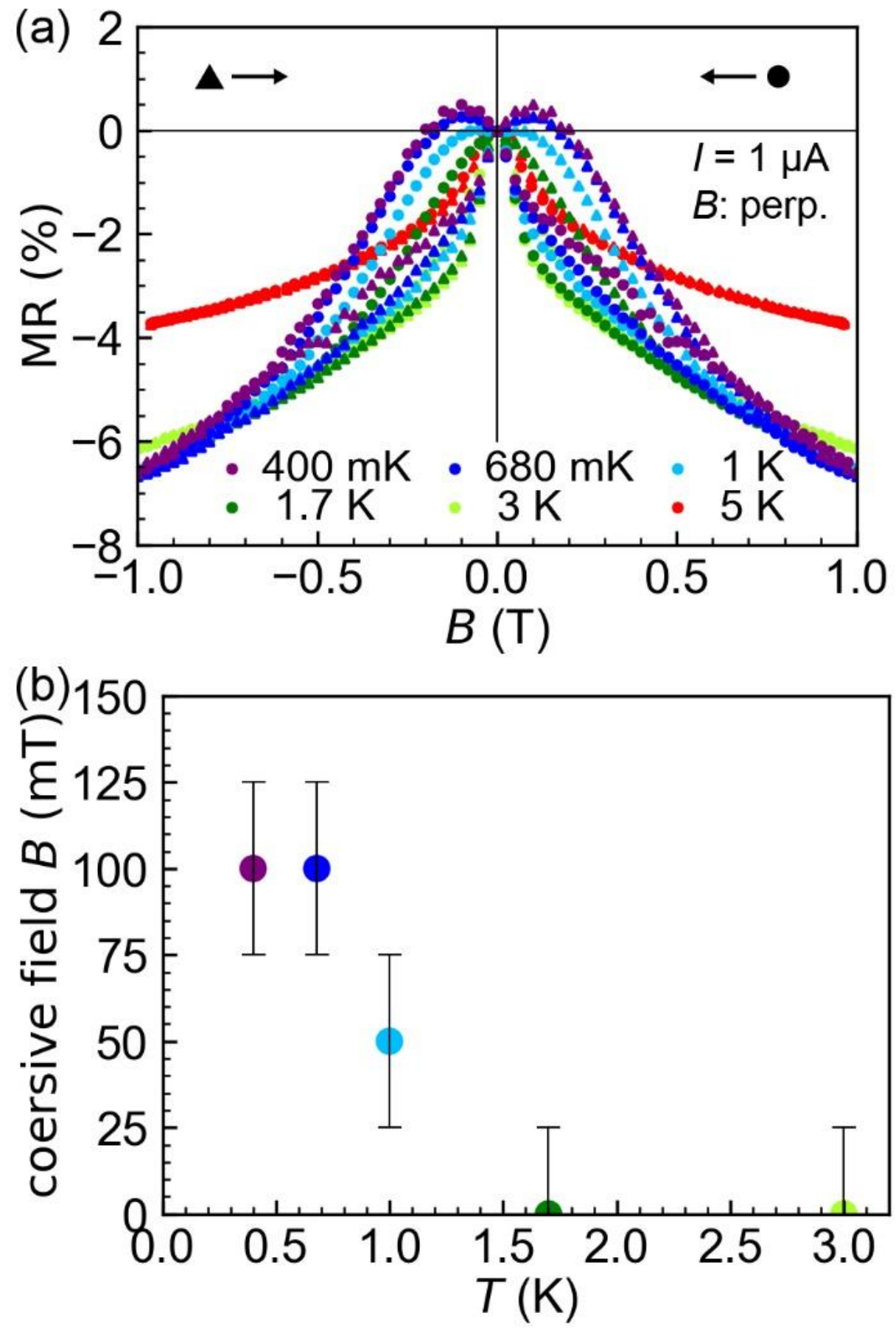

**Figure 4.** (a) Magnetoresistance of (In,Fe)As measured at different temperatures. The magnetic field is applied perpendicular to the sample plane, and the bias current is 1 μA. Triangle and circle symbols indicate magnetic field sweeps from negative to positive fields and from positive to negative fields, respectively. (b) Temperature dependence of the coercive field of (In,Fe)As, estimated from the peak positions in Fig. 4(a). The error bars correspond to the magnetic-field interval between adjacent measurement points.

# Supplementary Material

# Orientation selection and superconducting properties of epitaxial Al on ferromagnetic semiconductor (In,Fe)As

Hirotaka Hara[1], Keita Ishihara[1], Masaaki Tanaka[1,2,3,†], and Le Duc Anh[1,2,†]

[1] *Department of Electrical Engineering and Information System, The University of Tokyo, Tokyo, Japan*

[2] *Center for Spintronics Research Network, The University of Tokyo, Tokyo, Japan*

[3] *Institute for Nano Quantum Information Electronics, The University of Tokyo, Tokyo, Japan*

† Corresponding author: masaaki@ee.t.u-tokyo.ac.jp, anh@cryst.t.u-tokyo.ac.jp

## 1. Magnetic field dependence of the resistance of Al films at various temperatures

The magnetic-field dependence of the resistance ($R$) of Al was measured in the low-temperature regime, where Al is superconducting, as shown in Fig. S1. $B_{\text{in-plane}}$ is the in-plane magnetic field, $B_{\text{perp}}$ is the perpendicular magnetic field, and $I$ is the current. The superconducting state of Al is suppressed with increasing magnetic field and temperature. Here, $B_c$ is defined as the magnetic field at the intersection of linear-dashed-line fits to the zero-resistance region and the onset of non-zero resistance, which are shown by black dots in Fig. S1. These data are summarized as the temperature dependence of the critical magnetic field $B_c$ in Figs. 2(b) and 2(c) in the main manuscript.

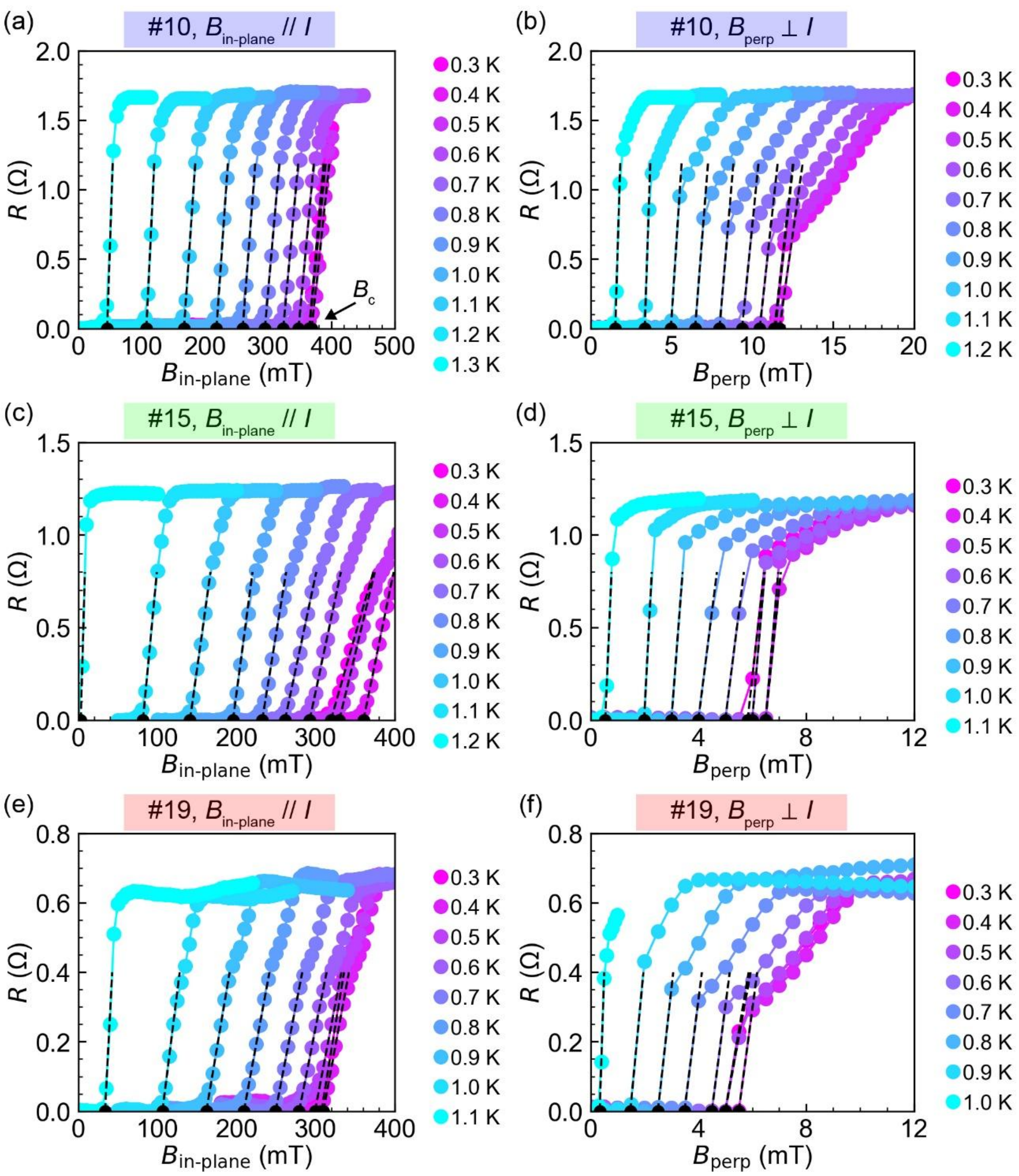

**Figure S1.** (a)-(f) Magnetic field dependence of the resistance ($R$) of Al in samples #10, #15, and #19, measured at low temperature where it exhibits superconductivity. $B_{\text{in-plane}}$ is the in-plane magnetic field, $B_{\text{perp}}$ is the perpendicular magnetic field, and $I$ is the current.

## 2. Magnetoresistivity of (In,Fe)As from room temperature to cryogenic temperatures

The magnetoresistance of (In,Fe)As was measured over a wider temperature range than that shown in Fig. 4(a). Figure S2(a) shows the raw magnetoresistance (magnetoresistivity) data, and Fig. S2(b) shows the normalized magnetoresistance defined as $\mathrm{MR}(B) = [R(B) - R(0)] / R(0)$, under a magnetic field applied perpendicular to the film plane. These data indicate that the magnetoresistance changes from positive to negative as temperature decreases from room temperature to low temperatures. The magnetoresistance becomes negative at 50 K, which is close to the Curie temperature of 40 K determined by magnetic circular dichroism (MCD). Hysteretic behavior in the magnetoresistance becomes clearly visible in the 5 K data and below. Although the magnetoresistance at 1 T is nearly saturated below 5 K, the peaks around 0 T split below 1 K, as discussed in the main text.

The enhancement of ferromagnetism at low temperature may be related to a temperature-dependent change in magnetic anisotropy, as observed in other ferromagnetic semiconductors [S1,S2], and/or to changes in the mechanism responsible for ferromagnetism in (In,Fe)As. In this material, Fe-rich regions formed by spinodal decomposition collectively contribute to ferromagnetism through carrier-mediated coupling in the conduction band [S3,S4]. The enhanced ferromagnetic behavior in (In,Fe)As at low temperatures may motivate further investigation of similar phenomena in other ferromagnetic semiconductors, including Fe-doped systems such as (Ga,Fe)Sb, (Al,Fe)Sb, and (In,Fe)Sb.

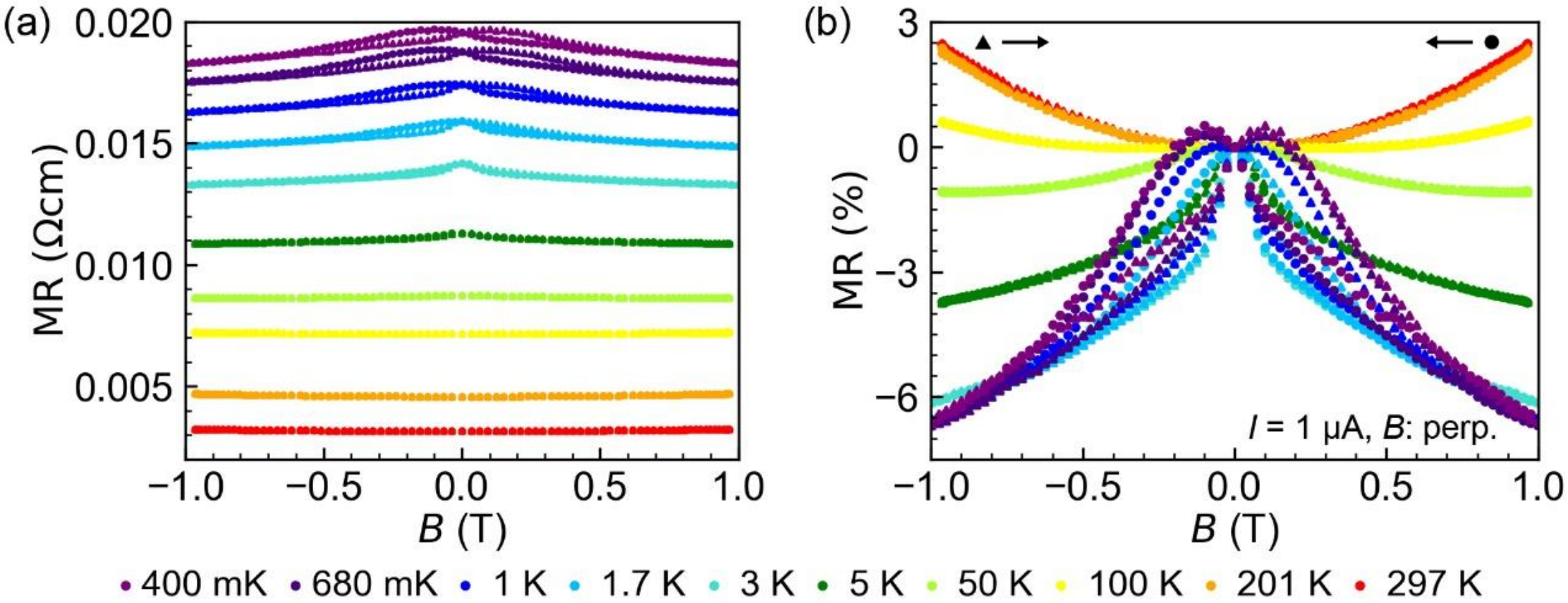

**Figure S2.** Temperature dependence of the magnetoresistivity of (In,Fe)As from 300 K to 400 mK, shown as (a) raw data and (b) normalized values, under a magnetic field perpendicular to the film plane**.**

### 3. Calculation of stray field generated by the (In,Fe)As layer

The stray field generated by the (In,Fe)As layer was calculated to evaluate its possible influence on the superconductivity in the adjacent Al layer. The calculation was performed at the center of the Al layer in a cleaved sample with typical dimensions used for electrical transport measurements, namely 8 mm × 4 mm × 15 nm for the (In,Fe)As layer. The magnetization $\boldsymbol{M}$ of (In,Fe)As was assumed to be uniaxial and oriented either perpendicular or in-plane, as shown in Figs. S3(a) and S3(b), respectively. The magnitude of the magnetization was based on the saturation magnetization of 40 emu/cm$^3$ reported in a previous study [S5]. The calculation was carried out using Maxwell's equations, following ref. [S6]. The stray fields generated by (In,Fe)As are calculated to be less than 1 μT for both perpendicular and in-plane magnetization, primarily because of the small magnetization of (In,Fe)As. The results indicate that the stray field is too small to account for the decrease in $B_c$ below 0.5 K, which corresponds to several mT to approximately 100 mT in Figs. 2(b) and 2(c).

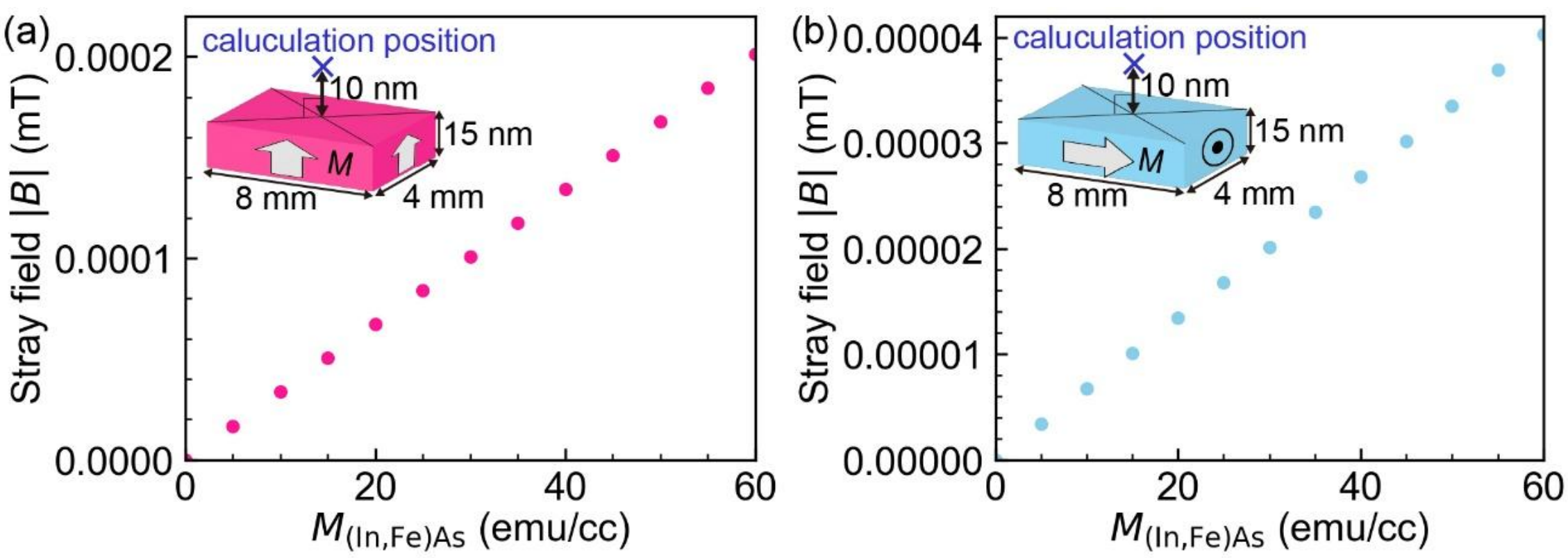

**Figure S3.** Dependence of the stray field on the magnetization of (In,Fe)As for (a) perpendicular magnetization and (b) in-plane magnetization.